\documentclass
[preprint,10pt,twocolumn,superscriptaddress,notitlepage,showpacs,nofootinbib,prl]{revtex4}%
\usepackage{amsfonts}
\usepackage{amsmath}
\usepackage{amssymb}
\usepackage{graphicx}%
\begin{document}
\title{Remarks about the Phase Transitions within the Microcanonical description}
\author{L. Velazquez}
\affiliation{Departamento de F\'{\i}sica, Universidad de Pinar del Rio, Marti 270, Esq. 27
de Noviembre, Pinar del Rio, Cuba}
\author{F. Guzman}
\affiliation{Instituto Superior de Tecnologia y Ciencias Aplicadas, Quinta de los Molinos,
Plaza de la Revolucion, La Habana, Cuba}

\begin{abstract}
According to the \textit{reparametrization invariance} of the microcanonical
ensemble, the only microcanonically relevant phase transitions are those
involving an \textit{ergodicity breaking} in the thermodynamic limit: the
zero-order phase transitions and the continuous phase transitions. We suggest
that the microcanonically relevant phase transitions are not associated
directly with topological changes in the configurational space as the
\textit{Topological Hypothesis} claims, instead, they could be related with
topological changes of certain subset $\mathcal{A}$ of the configurational
space in which the system dynamics is effectively trapped in the thermodynamic
limit $N\rightarrow\infty$.

\end{abstract}
\date{\today}
\maketitle

\section{Introduction}

Recently, a new characterization of phase transitions has been suggested by
Pettini and coworkers \cite{topH1,topH2,topH3,topH4}. According to the
\textit{Topological Hypothesis} proposed by these authors, there should be a
close relationship between the existence of a thermodynamic phase transition
at the macroscopic level and the existence of\textit{ changes in the
topological structure} of the configurational space of a generic many-body
Hamiltonian system:%
\begin{equation}
H\left(  q,p\right)  =\sum_{ij}\frac{1}{2}a^{ij}\left(  q\right)  p_{i}%
p_{j}+V\left(  q\right)  .
\end{equation}
A very important result obtained in this direction was the derivation of the
\textit{necessary character }of the topological changes for the existence of a
phase transition \cite{topH1,topH2,topH3,topH4}. The nowadays interest
concentrates in searching those \textit{sufficient and necessary conditions}
which lead to a topological classification scheme for phase-transitions.

As already shown in many studies \cite{topH1,topH2,topH3,topH4,app1}, most of
topological changes in the microscopic level do not provoke a phase
transition, apparently, only those strong topological changes. However,
Kastner have shown strong evidences about that a criterion based exclusively
on topological quantities cannot exist in general \cite{app7}. The efforts for
establishing the sufficient and necessary relations between topological
changes and phase transitions turn to be much more complicated by considering
the phenomenon of the ensemble inequivalence. The same author has shown some
evidences indicating that such close relation is expected to exist
\textit{only} between the topological approach and the microcanonical
characterization \cite{app8}.

We will show in the present Letter that the microcanonical description is
characterized by the existence of an internal symmetry: \textit{the
reparametrization invariance}. The presence of this symmetry implies a
revision of classification of phase transitions based on the concavity of the
Boltzmann entropy \cite{gro1}, as well as the question about the topological
origin of the phase transitions by starting from microcanonical basis.

\section{Reparametrization invariance}

Universality of the microscopic mechanisms of chaoticity provides a general
background for justifying the necessary ergodicity which supports a
thermostatistical description with microcanonical basis for all those
nonintegrable many-body Hamiltonian systems \cite{cohenG}. Thus, the
microcanonical ensemble:%
\begin{equation}
\hat{\omega}_{M}\left(  I,N\right)  =\frac{1}{\Omega\left(  I,N\right)
}\delta\left\langle I-\hat{I}\left(  X\right)  \right\rangle , \label{micro}%
\end{equation}
is just a \textit{dynamical ensemble} where every macroscopic characterization
has a direct mechanical interpretation. Here, $X$ represents a given point of
the phase space $\mathcal{X}$ and $\hat{I}\left(  X\right)  =\left\{  \hat
{I}^{1}\left(  X\right)  ,\hat{I}^{2}\left(  X\right)  ,\ldots\hat{I}%
^{n}\left(  X\right)  \right\}  $ are all those relevant (analytical)
integrals of motion determining the microcanonical description (generally
speaking, the\ total energy, the angular and linear momentum).

The admissible values of the set of integrals of motion $\hat{I}\left(
X\right)  $ could be considered as the "coordinate points" $I=\left\{
I^{1},I^{2},\ldots I^{n}\right\}  $\ of certain subset\ $\mathcal{R}_{I}$\ of
the n-dimensional Euclidean space $\mathcal{R}^{n}$. Each of these points
determines certain sub-manifold $\mathcal{S}_{p}$ of the phase space
$\mathcal{X}$ :%
\begin{equation}
X\in\mathcal{S}_{p}\equiv\left\{  X\in\mathcal{X}\left\vert \forall
k~I^{k}\left(  X\right)  =I^{k}\right.  \right\}  , \label{set}%
\end{equation}
in which the system trajectories spread uniformly in accordance with the
ergodic character of the microscopic dynamics. Such sub-manifolds defines a
partition $\Im$ of the phase space $\mathcal{X}$ in disjoint sub-manifolds:
\begin{equation}
\Im=\left\{  \mathcal{S}_{p}\subset\mathcal{X}\left\vert ~%
{\displaystyle\bigcup_{p}}
\mathcal{S}_{p}=\mathcal{X}~;~\mathcal{S}_{p}\cap\mathcal{S}_{q}%
=\varnothing\right.  \right\}  . \label{partition}%
\end{equation}
Definitions (\ref{set}) and (\ref{partition}) allow the existence of a
bijective map $\psi_{I}$ between the elements of $\Im$ (sub-manifolds
$\mathcal{S}_{p}\subset\mathcal{X}$) and the elements of $\mathcal{R}_{I}$
(points $I\in\mathcal{R}^{n}$):%

\begin{equation}
\psi_{I}:\Im\rightarrow\mathcal{R}_{I}\equiv\left\{  \forall\mathcal{S}_{p}%
\in\Im\left(  \mathcal{X}\right)  ~\exists I\in\mathcal{R}_{I}\subset
\mathcal{R}^{n}\right\}  .
\end{equation}
Thus, the partition $\Im$ has the same topological features of the
n-dimensional Euclidean subset $\mathcal{R}_{I}$. For this reason $\Im$ will
be referred as the \textit{abstract space of the integrals of motions}. We say
that the map $\psi_{I}$ defines the n-dimensional Euclidean \textit{coordinate
representation} $\mathcal{R}_{I}$ of the\ abstract space $\Im$.

Let us now to consider another subset $\mathcal{R}_{\varphi}\subset
\mathcal{R}^{n}$ with the same \textit{diffeomorphic structure} of the subset
$\mathcal{R}_{I}$ and the following diffeormorphic map $\varphi$ among them:%
\begin{equation}
\varphi:\mathcal{R}_{I}\rightarrow\mathcal{R}_{\varphi}\equiv\left\{  \forall
I\in\mathcal{R}_{I}~\exists\varphi\in\mathcal{R}_{\varphi}\left\vert
\det\left(  \frac{\partial\varphi^{j}}{\partial I^{k}}\right)  \not =0\right.
\right\}  . \label{diffeomorphic}%
\end{equation}
We say that the map $\varphi$ represents a general \textit{reparametrization
change} of the microcanonical description since it allows us to introduce
another n-dimensional Euclidean coordinate representation $\mathcal{R}%
_{\varphi}$ by considering the bijective map $\psi_{\varphi}=\psi_{I}%
o\varphi^{-1}$:%
\begin{equation}
\psi_{\varphi}:\Im\rightarrow\mathcal{R}_{\varphi}\equiv\left\{
\forall\mathcal{S}_{p}\in\Im~\exists\varphi\in\mathcal{R}_{\varphi}%
\subset\mathcal{R}^{n}\right\}  .
\end{equation}
The above reparametrization change $\varphi$ also induces the following
reparametrization of the relevant integrals of motion $\varphi_{X}:\hat
{I}\left(  X\right)  \rightarrow\hat{\varphi}\left(  X\right)  $, where:
\begin{equation}
\hat{\varphi}\left(  X\right)  =\left\{  \varphi^{1}\left\langle \hat
{I}\left(  X\right)  \right\rangle ,\varphi^{2}\left\langle \hat{I}\left(
X\right)  \right\rangle ,\ldots\varphi^{n}\left\langle \hat{I}\left(
X\right)  \right\rangle \right\}  .
\end{equation}
Since $\hat{I}\left(  X\right)  $ are integrals of motions, every $\varphi
^{k}\left\langle \hat{I}\left(  X\right)  \right\rangle \in\hat{\varphi
}\left(  X\right)  $ will be also an integral of motion. The bijective
character of the reparametrization change $\varphi:\mathcal{R}_{I}%
\rightarrow\mathcal{R}_{\varphi}$ allows us to say that the sets $\hat
{\varphi}\left(  X\right)  $ and $\hat{I}\left(  X\right)  $ are
\textit{equivalent representations} of the relevant integrals of motion of the
microcanonical description because of they generate the same phase space
partition $\Im$ (\ref{partition}).

The interesting question is that the microcanonical ensemble is
\textit{invariant} under every reparametrization change. Considering the
identity:%
\begin{equation}
\delta\left\langle \varphi-\hat{\varphi}\left(  X\right)  \right\rangle
\equiv\left\vert \frac{\partial\varphi}{\partial I}\right\vert ^{-1}%
\delta\left\langle I-\hat{I}\left(  X\right)  \right\rangle ,
\end{equation}
where $\left\vert \partial\varphi/\partial I\right\vert \not =0$ is the
Jacobian of the reparametrization change $\varphi$, the phase space
integration leads to the following transformation rule for the microcanonical
partition function:%
\begin{equation}
\Omega\left(  \varphi,N\right)  =\left\vert \frac{\partial\varphi}{\partial
I}\right\vert ^{-1}\Omega\left(  I,N\right)  ,
\end{equation}
leading in this way to the reparametrization invariance of the microcanonical
distribution function:%
\begin{equation}
\frac{1}{\Omega\left(  \varphi,N\right)  }\delta\left\langle \varphi
-\hat{\varphi}\left(  X\right)  \right\rangle \equiv\frac{1}{\Omega\left(
I,N\right)  }\delta\left\langle I-\hat{I}\left(  X\right)  \right\rangle .
\label{invariance}%
\end{equation}

A corollary of the identity (\ref{invariance}) is that the Physics derived
from the microcanonical description is reparametrization invariant since the
expectation values of every macroscopic observable $\hat{O}\left(  X\right)  $
obtained from the microcanonical distribution function $\hat{\omega}%
_{M}\left(  X\right)  $ exhibits this kind of symmetry:%
\begin{equation}
\bar{O}=\int\hat{O}\left(  X\right)  \hat{\omega}_{M}\left(  X\right)
dX\Rightarrow\bar{O}\left(  \varphi,N\right)  =\bar{O}\left(  I,N\right)  .
\end{equation}
The reparametrization invariance does not introduce anything new in the
macroscopic description of a given system, except the possibility of
describing the microcanonical macroscopic state by using any coordinate
representation of the abstract space $\Im$, a situation analogue to the
possibility of describing the physical space $\mathcal{R}^{3}$ by using a
Cartesian coordinates $\left(  x,y,z\right)  $ or a spherical coordinates
$\left(  r,\theta,\varphi\right)  $. Thus, we can develop a geometrical
formulation of Thermostatistics within the microcanonical ensemble.

The microcanonical partition function allows us to introduce an invariant
measure $d\mu=\Omega dI$ for the abstract space $\Im$, leading in this way to
an invariant definition of the Boltzmann entropy $S_{B}=\ln W$, where
$W=\int_{\Sigma_{\alpha}}d\mu$ \ characterizes certain \textit{coarse grained}
partition $\left\{  \Sigma_{\alpha}\left\vert ~%
{\textstyle\bigcup\nolimits_{\alpha}}
\Sigma_{\alpha}=\Im\right.  \right\}  $. In the thermodynamic limit
$N\rightarrow\infty$ the coarsed grained nature of the Boltzmann entropy can
be disregarded and taken as a scalar function defined on the space $\Im$.

\section{Microcanonical Phase Transitions}

The present proposal comes from by arising the above reparametrization
invariance to a fundamental status within the microcanonical description. In
the sake of simplicity, let us consider a Hamiltonian system with a
microscopic dynamics driven by short-range forces, so that, it becomes
extensive in the thermodynamic limit. Let us suppose also that its
microcanonical description is determined from the consideration of only one
integral of motion: the total energy $E$. Among all different
reparametrizations of the total energy $E$ which can be taken into account, we
shall limit only to the following generic form $\Theta=N\varphi\left(
E/N\right)  $, where $N$ represents the system size and $\varphi\left(
\varepsilon\right)  $, an analytical bijective function of the energy per
particle $\varepsilon=E/N$. Obviously, the quantity $\Theta$ represents an
integral of motion of the microscopic dynamics which has the advantage of
preserving the same extensive character of the energy in the thermodynamic
limit. Hereafter, the function $\varepsilon\left(  \varphi\right)  $
represents the inverse of the function $\varphi\left(  \varepsilon\right)  $.

Taking into account the dynamical origin of the microcanonical description, a
phase transition within this ensemble should be the macroscopic manifestation
of certain sudden change in the microscopic level which manifests itself as a
mathematical anomaly of the Boltzmann entropy. Since the entropy is ussually
an analytical function in a finite system, the most important mathematical
anomalies of the entropy per particle $s\left(  \varphi\right)  =S_{B}%
\left\langle \varepsilon\left(  \varphi\right)  N,N\right\rangle /N$ are
(\textbf{A}) the existence of regions where this function is not locally
concave (convex down), $\partial^{2}s\left(  \varphi\right)  /\partial
\varphi^{2}\geq0$, as well as (\textbf{B}) every lost of analyticity in the
thermodynamic limit $N\rightarrow\infty$.

The\ non concavity of the entropy is usually related with the phenomenon of
ensemble inequivalence between the microcanonical description and the one
performed by using the Gibbs canonical ensemble, which is associated with the
occurrence of the first order phase transitions. However, the behavior
\textbf{A} represents an anomaly within the canonical description because of
there in nothing anomalous within the microcanonical ensemble: these regions
represent microcanonical thermodynamic states with a negative heat capacity,
which can not be accessed within the canonical ensemble when the thermodynamic
limit is invoked \cite{gro1}. A negative heat capacity in systems with
short-range interactions outside the thermodynamic limit is identify with the
existence of a non-vanishing interphase surface tension \cite{gro1}. While
this phenomenon disappears in these systems with the imposition of the
thermodynamic limit, it survives in systems with long-range interactions, i.e.
the astrophysical systems \cite{ant,chava}. Although they are non-homogeneous,
a negative heat can not be always identified with the existence of interphase
boundaries, which can be verified by reexamining the Antonov isothermal model
\cite{ant}.

The reader can notice by considering the microcanonical reparametrization
invariance that the convex up or down character of any scalar function is
\textit{ambiguous}: it depends on the coordinate representation used for
describe it. Let us see a trivial example. Let $s$ be a positive real map
defined on a seminfinite Euclidean line $\mathcal{R}_{1}$, $s:\mathcal{R}%
_{1}\rightarrow\mathcal{R}^{+}$, which is given by the concave function
$s\left(  x\right)  =\sqrt{x}$ in the coordinate representation $\mathcal{R}%
_{x}$ of $\mathcal{R}_{1}$ (where $x>0$). Let $\varphi$ be a reparametrization
change $\varphi:\mathcal{R}_{x}\rightarrow\mathcal{R}_{y}$ given by
$y=\varphi\left(  x\right)  =x^{\frac{1}{4}}$ (which is evidently a bijective
map). The map $s$ in the new representation $\mathcal{R}_{y}$ is given by the
function $s\left(  y\right)  =y^{2}$ (where $y>0$), which clearly is a convex
function in this coordinate representation of the domain $\mathcal{R}_{1}$.

Since the convexity of the Boltzmann entropy depends crucially on the
reparametrization, the ensemble inequivalence between the microcanonical
description and the one performed by using the following
\textit{generalization }of the Gibbs canonical ensemble:
\begin{equation}
\hat{\omega}_{c}\left(  \eta,N\right)  =\frac{1}{Z\left(  \eta,N\right)  }%
\exp\left\langle -\eta\Theta\right\rangle \label{canonical}%
\end{equation}
depends also on the reparametrization $\Theta=N\varphi\left(  E/N\right)  $.
Such noninvariance of the ensemble inequivalence follows from the fact that
the distribution function (\ref{canonical}) does not obey the original
reparametrization invariance since this ensemble does not describe an
\textit{isolate} Hamiltonian system: the coordinate representation $\Theta$
used in the canonical description has been determined from certain external
constrains which has been imposed to the interest system. This idea is very
easy to understand by analyzing the case of the extensive systems: the
canonical ensemble $\omega_{BG}=Z^{-1}\left(  \beta,N\right)  \exp\left(
-\beta H_{N}\right)  $ is experimentally implemented by putting the interest
system in thermal contact with a heat bath. This experimental arrangement
keeps fixed not only the system temperature $T=\beta^{-1}$, but also the
coordinate representation by using the system total energy, $\Theta\equiv$
$E$. An arbitrary reparametrization change $E\rightarrow\Theta$ within the
canonical ensemble (\ref{canonical}) is physically implemented by considering
another experimental arrangement which keeps fixed the canonical parameter
$\eta=\partial s\left(  \varphi\right)  /\partial\varphi$. The possibility of
using different reparametrizations $\Theta$ in the canonical distribution
function allows us to avoid the ensemble inequivalence in those thermodynamic
states with a negative heat capacity, a feature particularly useful for
enhancing the possibilities of some general Monte Carlo methods inspired on
the Statistical Mechanics. This idea was applied in ref.\cite{vel-mmc} to
improve the well-known Metropolis importance sampling algorithm \cite{met},
which is usually unable to describe the thermodynamical states with a negative
heat capacity.

Although paradoxical, the identification of the first order phase transitions
with the ensemble inequivalence allows us to claim that this kind of phase
transitions are not microcanonically relevant because of they are irrelevant
from the dynamical viewpoint. Contrary, it is very easy to verify that every
loss of analyticity of the microcanonical entropy in the thermodynamic limit
appears without mattering about the analytical function $\varphi\left(
\varepsilon\right)  $ used in the reparametrization $\Theta$. Thus, the
mathematical anomaly \textbf{B} is compatible with the reparametrization
invariance of the microcanonical ensemble, and it is apparently the
macroscopic manifestation of a sudden change in the behavior of the
microscopic dynamics of the system. An reexamination of the available
experimental and theoretical results suggests us a direct connection of
anomaly \textbf{B} with the occurrence of an \textit{ergodicity breaking}.
Ergodicity breaking takes place when the time averages and the ensemble
averages of certain macroscopic observables can not be identified due to the
microscopic dynamics is effectively trapped in different subsets of the
configurational or phase space during the imposition of the thermodynamic
limit $N\rightarrow\infty$ \cite{Gold}. Let us see two examples.

It is well-known that ensemble equivalence holds during the \textit{continuous
phase transitions} in systems with short-range interactions, but the heat
capacity $c\left(  \varepsilon\right)  =-\beta^{2}\left(  \varepsilon\right)
/\kappa\left(  \varepsilon\right)  $ diverges at the critical energy
$\varepsilon_{c}$ in the thermodynamic limit where $\kappa\left(
\varepsilon_{c}\right)  =0$, being $\beta=\partial s\left(  \varepsilon
\right)  /\partial\varepsilon$ and $\kappa\left(  \varepsilon\right)
=\partial^{2}s\left(  \varepsilon\right)  /\partial\varepsilon^{2}$. It is not
difficult to show by using the Taylor power series expansion of the caloric
curve $\beta\left(  \varepsilon\right)  $ that the analyticity of the entropy
$s\left(  \varepsilon\right)  $ at the critical energy is unable to explain
the existence of nontrivial critical exponent $\alpha$ in the heat capacity
close to the critical inverse temperature\ $\beta_{c}=\beta\left(
\varepsilon_{c}\right)  $:%
\begin{equation}
c\left(  \beta\right)  \simeq\left\{
\begin{array}
[c]{cc}%
A^{-}/\left(  \beta_{c}-\beta\right)  ^{\alpha} & \text{if }\beta<\beta_{c},\\
A^{+}/\left(  \beta-\beta_{c}\right)  ^{\alpha} & \text{if }\beta>\beta_{c},
\end{array}
\right.
\end{equation}
with a universal ratio $A^{-}/A^{+}\not =1$ \cite{Gold}. This non-analyticity
of the entropy in the thermodynamic limit is clearly associated with the
occurrence of an ergodicity breaking during the continuous phase transitions
as a consequence of an underlying symmetry breaking.

Another evidence of such connection is during the occurrence of the called
\textit{zero-order phase transitions} in systems with long-range interactions
\cite{chava}. This anomaly manifests itself as a discontinuity in the first
derivative of the entropy, which represents the existence of equiprobable
metastable states with different temperatures at the same total energy.
Depending from the initial conditions, only one of these metastable
configurations will be given in practice when $N\rightarrow\infty$.

Apparently, every lost of analyticity of the entropy in the thermodynamic
limit can be associated with the existence of several metastable states
(sub-manifolds in the configurational space) in which the system dynamics can
be effectively trapped in the thermodynamic limit: While all these metastable
states in the continuous phase transitions are related by an internal symmetry
of the microscopic dynamics, the metastable states in the zero-order phase
transitions are essentially different (not related by an internal symmetry).
These are the only phase transitions which are relevant anomalies within the
microcanonical ensemble. They have an evident dynamical origin which can be
associated with mathematical anomalies of the entropy compatible with the
reparametrization invariance of the microcanonical ensemble.

\textit{Does every phase transition have a Topological origin?} We think that
the answer to this question is negative due to the only phase transitions
which are dynamically relevant in an isolate nonintegrable many-body
Hamiltonian system are those involving an ergodicity breaking in the
microscopic picture. While a classification scheme of the phase transitions
based exclusively on the topology of configurational space cannot exist in
general \cite{app7}, it is not difficult to understand that an ergodicity
breaking in the thermodynamic limit always involves certain \textit{effective
topological change} in the configurational space.

Let us consider a classical 2-dimensional ferromagnetic model system with a
microscopic magnetization density $\mathbf{m}$ given by $\mathbf{m}\left[
\theta\right]  =N^{-1}\sum_{i}\left(  \cos\theta_{i},\sin\theta_{i}\right)  $,
whose Hamiltonian exhibits a $U\left(  1\right)  $ symmetry, and where the
configurational space is the \textit{N}-dimensional tori $\mathcal{C}=\left\{
0<\theta_{i}\leq2\pi;~i=1,..N\right\}  $. The equilibrium distribution
function of the modulus of the magnetization density $m=\left\vert
\mathbf{m}\right\vert $ when $N$ is large enough exhibits a very sharp
Gaussian profile around the certain average value $m_{0}\left(  \varepsilon
\right)  $ which depends on the energy per particle $\varepsilon$, with a
dispersion decreasing as $\sigma_{m}\propto1/\sqrt{N}$. Thus, the imposition
of the thermodynamic limit leads to an \textit{effective trapping} of the
microscopic dynamics in the following subset of the configurational space:
\begin{equation}
\mathcal{A}\left(  \varepsilon\right)  =\left\{  \theta\in\mathcal{C}%
\left\vert ~\left\vert \mathbf{m}\left[  \theta\right]  \right\vert
=m_{0}\left(  \varepsilon\right)  \right.  \right\}  .
\end{equation}
Since the average magnetization density vanishes identically in the
paramagnetic phase (with $\varepsilon$ greater than certain critical energy
$\varepsilon_{c}$), the dimension of the subset $\mathcal{A}\left(
\varepsilon\right)  $ is $\dim\left\langle \mathcal{A}\left(  \varepsilon
\right)  \right\rangle =N-2$, and its codimension $C\left\langle
\mathcal{A}\left(  \varepsilon\right)  \right\rangle =\dim\mathcal{C}%
-\dim\left\langle \mathcal{A}\left(  \varepsilon\right)  \right\rangle =2$ is
topological invariant in the thermodynamic limit $N\rightarrow\infty$. In the
ferromagnetic phase with $\varepsilon<\varepsilon_{c}$, the system exhibits a
spontaneous magnetization with an arbitrary orientation due to the $U(1)$
symmetry. The dimension of the subset $\mathcal{A}\left(  \varepsilon\right)
$ is now $\dim\left\langle \mathcal{A}\left(  \varepsilon\right)
\right\rangle =N-1$, and its codimension is given by $C\left\langle
\mathcal{A}\left(  \varepsilon\right)  \right\rangle =\dim\mathcal{C}%
-\dim\left\langle \mathcal{A}\left(  \varepsilon\right)  \right\rangle =1$,
which is also topological invariant when $N\rightarrow\infty$. Thus, during
the continuous phase transition there is a topological change of the
codimension $C\left\langle \mathcal{A}\left(  \varepsilon\right)
\right\rangle $ of the subset $\mathcal{A}\left(  \varepsilon\right)  $.

This example suggests us that the microcanonically relevant phase transitions
are not directly associated with topological changes in the configurational
space as the \textit{Topological Hypothesis} claims
\cite{topH1,topH2,topH3,topH4}, instead, we think that they could be related
with certain topological change of subset $\mathcal{A}$ of the configurational
space in which the system dynamics is effectively trapped in the thermodynamic
limit $N\rightarrow\infty$.

\end{document}